\documentclass{aa}

\usepackage{graphicx}
\usepackage{subfigure}
\usepackage{txfonts}
\usepackage{gensymb} 
\usepackage{stfloats}
\usepackage[table]{xcolor}
\usepackage{float}
\usepackage{hyperref}
\usepackage{placeins}
\usepackage{afterpage}

\definecolor{mycolor}{RGB}{255,200,200}
\begin{document}

   \title{CAPOS: the bulge Cluster APOgee Survey IX. Spectroscopic Analysis of the Bulge Globular Cluster Terzan 2}


   \author{Macarena Uribe
          \inst{1}
          \and
          Sandro Villanova\inst{2}\fnmsep\thanks{Just to show the usage
          of the elements in the author field}
          \and
          Douglas Geisler\inst{1,3}
          \and José G. Fernández-Trincado\inst{4}
          \and Cesar Muñoz\inst{3}
          \and Nicolás Barrera\inst{3}
          \and Thaiz Pino\inst{1}
          \and Franco Sepúlveda\inst{1}
          }

   \institute{Departamento de Astronomía, Casilla 160-C, Universidad de Concepción, Concepción, Chile
   \and{Universidad Andres Bello, Facultad de Ciencias Exactas, Departamento de F{\'i}sica y Astronom{\'i}a - Instituto de Astrof{\'i}sica, Autopista Concepci\'on-Talcahuano 7100, Talcahuano, Chile}
    \and Departamento de Astronomía, Facultad de Ciencias, Universidad de La Serena, Av. Juan Cisternas 1200, La Serena, Chile
    \and Universidad Cat\'olica del Norte, N\'ucleo UCN en Arqueolog\'ia Gal\'actica - Inst. de Astronom\'ia, Av. Angamos 0610, Antofagasta, Chile, jose.fernandez@ucn.cl
             }

   \date{Received XXXXX; accepted XXXXX}

 
  \abstract
   {We present the first detailed spectroscopic analysis of the heavily extincted bulge globular cluster Terzan 2 (Ter 2) based on high-resolution near-infrared spectra obtained as part of the CAPOS (the bulge Cluster APOgee Survey) project. CAPOS focuses on surveying clusters within the Galactic Bulge, using the APOGEE-2S spectrograph, part of the SDSS-IV survey, a component of the second generation Apache Point Observatory Galactic Evolution Experiment (APOGEE-2). For the spectral analysis, we use the Brussels Automatic Code for Characterizing High accUracy Spectra (BACCHUS) code which provides line-by-line elemental abundances. We derive abundances for Fe-peak (Fe, Ni), $\alpha$-(O, Mg, Si, Ca, Ti), light-(C, N), odd-Z(Al), and the s-process element (Ce) for four members of the cluster. Our analysis yields a mean metallicity of [Fe/H]= $-$0.84 $\pm$ 0.04, with  no evidence for an intrinsic variation. We detect a significant abundance variation only in C and N, indicating the presence of multiple populations. Ter 2 exhibits a typical $\alpha$-enrichment, that follows the trend of Galactic GCs. Additionally, our dynamical analysis reveals that Terzan 2 is a bulge globular cluster with a chaotic orbit that is influenced by the Galactic bar but not trapped by it, displaying both prograde and retrograde motions within the inner bulge region. Overall, the chemical patterns observed in Terzan 2 are in good agreement with those of other CAPOS bulge clusters of similar metallicity.}


\titlerunning{CAPOS: the bulge Cluster APOgee Survey IX}
\authorrunning{Macarena Uribe et al.}
   \maketitle
%

\section{Introduction}

Globular clusters (GCs) located in the Bulge of the Milky Way are thought to be relics of the early epoch of formation and evolution of the oldest structure in the Galaxy \citep{bica2016,barbuy18,perezvillega20}. Despite this, Bulge GCs (BGCs) have been poorly studied so far \citep{recioblanco17,ftrincado19}. This is due to the large amount of dust in the Galactic plane and towards the bulge, which generates a high and often variable extinction, making optical observations of these objects difficult \citep{nataf19}. In addition, all major components of the Galaxy - bulge, thin and thick disks, and halo - reach their highest densities in the inner Galaxy, making both crowding as well as contamination serious problems in areas where extinction is less of an issue. 

On the other hand, our understanding of the formation and evolution of GCs has changed dramatically since observations have found the presence of multiple populations (MPs) in them, via the the identification of star-to-star variations in the light- (C,N), odd-Z (Na, Al), $\alpha$-(O, Mg) as well as $s$-elements \citep{ftrincado21sep}. These internal variations demonstrate the complex enrichment in which GCs must have formed. This also provides clues into the conditions of the early Universe in which these systems formed \citep{renzini15}. The presence of MPs implies an initial population of stars formed from gas enriched by Type II supernovae, thus with an overabundance of alpha elements, and a second population of stars presumably formed subsequently from the remaining gas polluted by massive stars of the first population, as reflected in the C-N, Na-O and Mg-Al anti-correlations \citep{smith87,meszaros15,caretta09}. However, the nature of the first-population polluters (e.g. fast rotating massive stars, massive binaries, super massive stars and/or intermediate mass AGB stars) and the exact mechanism of the formation of the second population remain unknown\citep{bastianlardo18}. Further observations are demanded to help constrain various scenarios

The problem of high extinction is largely minimized with the use of near-infrared (NIR) photometry and high-resolution spectroscopy, which make it possible to penetrate the obscuring dust. In this way, a detailed mapping of the chemical and kinematic properties of heavily optically-extincted GCs has been possible. In this context, the Apache Point Observatory Galactic Evolution Experiment (APOGEE, \citet{majewski17}) of the Sloan Digital Sky Survey-IV provided the first high-resolution (R$\sim$22500) near-IR (H-band) multiobject spectrograph in the south. An obvious target are the large sample of bulge GCs. However, the SDSS-IV APOGEE survey did not prioritize such objects despite the rather large sample of fields covered in the bulge. Recognizing this opportunity, CAPOS (the bulge Cluster APOgee Survey) was undertaken as a Contributed program to SDSS-IV \cite{geisler21}. A major goal was to obtain high resolution spectra for a number of members in each of a number of BGCs in order to study their basic chemistry via mean abundances. In addition, CAPOS opened up the study of metal-rich GCs in order to expand the parameter space to explore MPs, since the most metal-rich GCs are only found in the bulge or disk  (\cite{bica24}, \cite{geisler25}). CAPOS also benefits from the high-precision astrometric and photometric data from the latest release of the ESA Gaia mission (Gaia DR3, \cite{gaia23}), combined with the photometric depth of the Two-Micron All Sky Survey \citep{skurtskie06} and the infrared VISTA Variables in the Vía Láctea (VVV, \citet{minniti10} ; \citet{saito12}) survey. In the end, CAPOS observed a total of 18 GCs located towards the bulge. An overview of CAPOS and initial results for a subsample were presented in \citet{geisler21}, based on the APOGEE Stellar Parameters and Chemical Abundance Pipeline (ASPCAP, \citep{garciaperez16}) incorporated in SDSS-IV. A second study, focused on FSR 1758, was published by \citet{rcolmenares21}. The third work corresponds to the first high-resolution spectroscopic analysis of Ton 2 \citep{ftrincado22}. Subsequent studies include NGC 6558 \citep{gonzalezdiaz23}, HP 1 \citep{lady25}, NGC 6569 \citep{barrera25} NGC 6316 \cite{frelij25} and the chemical composition of Dorj 2 \citep{pino25}. All of these latter studies focussed on a single cluster and employed the  BACCHUS package (Brussels Automatic Code for Characterizing accUracy Spectra; \cite{masseron16}. Most recently, \cite{geisler25} present final results based on ASPCAP for all CAPOS GCs. This paper is the tenth in the CAPOS series.

\begin{figure}[h]
    \centering
    \includegraphics[width=0.45\textwidth]{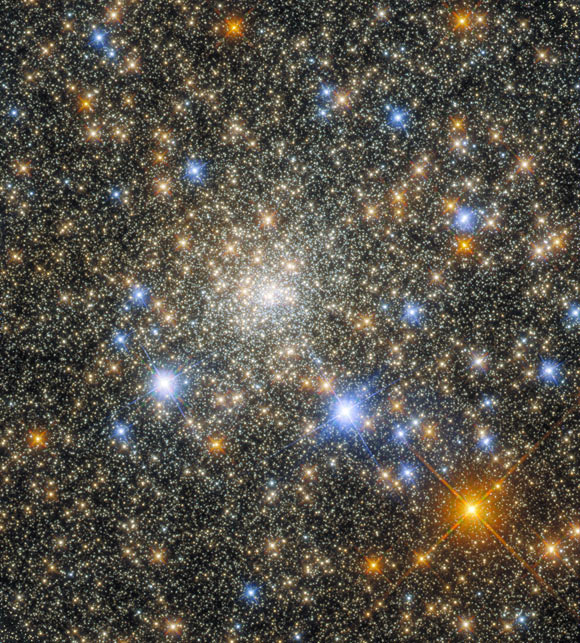} 
    \caption{The bulge globular cluster Terzan 2 capture by the Hubble Space Telescope. | Credit: NASA, ESA, ESA/Hubble, Roger Cohen (RU)}
    \label{fig:mi_imagen}
\end{figure}

The object of this study, which we can see in Figure \ref{fig:mi_imagen}, was first cataloged as Terzan 2.  It is a GC located towards the bulge, with $l$= 356.3, $b$= 2.30. It has a very high reddening, estimated by \cite{harris10} as E(B-V)=1.87. It has been the subject of multiple studies aimed at constraining its fundamental parameters, particularly its metallicity. Early measurements based on integrated infrared photometry estimated [Fe/H]$\approx -0.47$ \citep{malkan82}, while a later study suggested a slightly higher metallicity of [Fe/H]$\approx -0.25$ from integrated spectroscopy and infrared color-magnitude diagrams \citep{kuchinski95}. Terzan 2 is a globular cluster located in the bulge of the Milky Way. It is characterized by an intermediate metallicity for a BGC, with a recently reported value of [Fe/H] = $-$0.85 (\cite{geisler21}) while \cite{harris10} gives $-$0.69.

According to \cite{Baumgardt2021} the cluster has a mass of $8.05\pm2.31\cdot10^{4} M_{\odot}$, a distance from the Sun of $7.75\pm0.33$ kpc, a galactocentric distance of $0.74\pm0.16$ kpc and a tidal radius of 12.45 pc. This BGC is an important object for studying the chemical and dynamical evolution of the inner regions of our galaxy, offering insights into the complex formation history of the bulge and its stellar populations.

CAPOS data provide a spectroscopic metallicity more robust than any current available for Terzan 2, as well as detailed abundances for many elements with a wide variety of nucleosynthetic origins, incluing the $\alpha$-(0, Mg, Si, Ca, Ti), Fe-peak (Fe, Ni), light-(C, N), odd-Z(Al) and $s$-process (Ce) elements. The chemical abundances of Terzan 2 were derived using the BACCHUS package (Brussels Automatic Code for Characterizing accUracy Spectra; \citealt{masseron16}).A rigorous line-by-line inspection was performed across spectra encompassing a range of abundance variations, allowing for the determination of optimal fit values for each element with maximal reliability.

This paper is organized as follows: Section \ref{sec:the data} describes the spectroscopic data used. In Section \ref{sec:target selection} we review the target selection. Section \ref{sec:atmospheric parameters} shows the parameters selection. In Section \ref{sec:abundances} we present the abundance determinations. The results are described in Section \ref{sec:results}. A dynamical analysis of the cluster is present in Section \ref{sec:dynamical history} and, finally, we present our summary and concluding remarks in Section \ref{sec:conclusions}.

\section{Spectroscopic data}
\label{sec:the data}
The analysis was carried out on high-resolution (R$\approx$ 22,500) near-infrared spectra obtained by the Apache Point Observatory Galactic Evolution Experiment II survey (APOGEE-2; \citet{majewski17}), an internal program of SDSS-IV \citep{blanton17}, with the aim of delivering accurate radial velocities and detailed chemical abundances for a large sample of giant stars covering all components of the Milky Way and nearby dwarf companions.\\
The APOGEE-2 survey uses two spectrographs \citep{wilson19} one from the northern hemisphere on the 2.5 m telescope at Apache Point Observatory (APO, APOGEE-2N;\citet{gunn06}) and another from the southern hemisphere on the Irénée du Pont 2.5 m telescope \citep{bowenandv73} at Las Campanas Observatory (LCO, APOGEE-2S). Each instrument records most of the \textit{H} band (1.5 $\mu$m $-$ 1.69 $\mu$m) on three detectors, with coverage gaps between $\sim$ 1.58$-$1.59$\mu$m and $\sim$1.64$-$1.65$\mu$m, and with each fiber subtending 2" in diameter on the sky in the northern instrument and 1.3" in the southern. 

The final release of APOGEE-2, Data Release 17 (DR17), from SDSS-III/SDSS-IV, includes all data collected at APO through November 2020 and at LCO through January 2021, with the two APOGEE-2 instruments having observed over 650,000 stars across the Milky Way. Target selection is detailed in \citet{zasowski17}, \citet{beaton21} and \citet{santana21}. Spectra were reduced as described in \citet{nidever15}, and analyzed using the APOGEE Stellar Parameters and Chemical Abundance Pipeline (ASPCAP; \citet{garciaperez16}), and the libraries of synthetic spectra described in \citet{zamora15}. The customized \textit{H}-band line lists are fully described in \citet{shetrone15}, \citet{hasselquist16} (neodymium lines (Nd II)), \citet{cunha17} (cerium lines (Ce II)) and \citet{smith21}.
CAPOS was a Contributed APOGEE program (Geisler PI) allocated by the CNTAC designed to study as large a sample of BGCs. Most recently, \cite{geisler25} present final ASPCAP-based abundances for the entire 18 cluster CAPOS sample. Terzan 2 was observed as part of the contributing CAPOS survey \citep{geisler21}.

\begin{figure*}[!ht]
    \centering
    \subfigure[]{\includegraphics[height=0.46\textwidth]{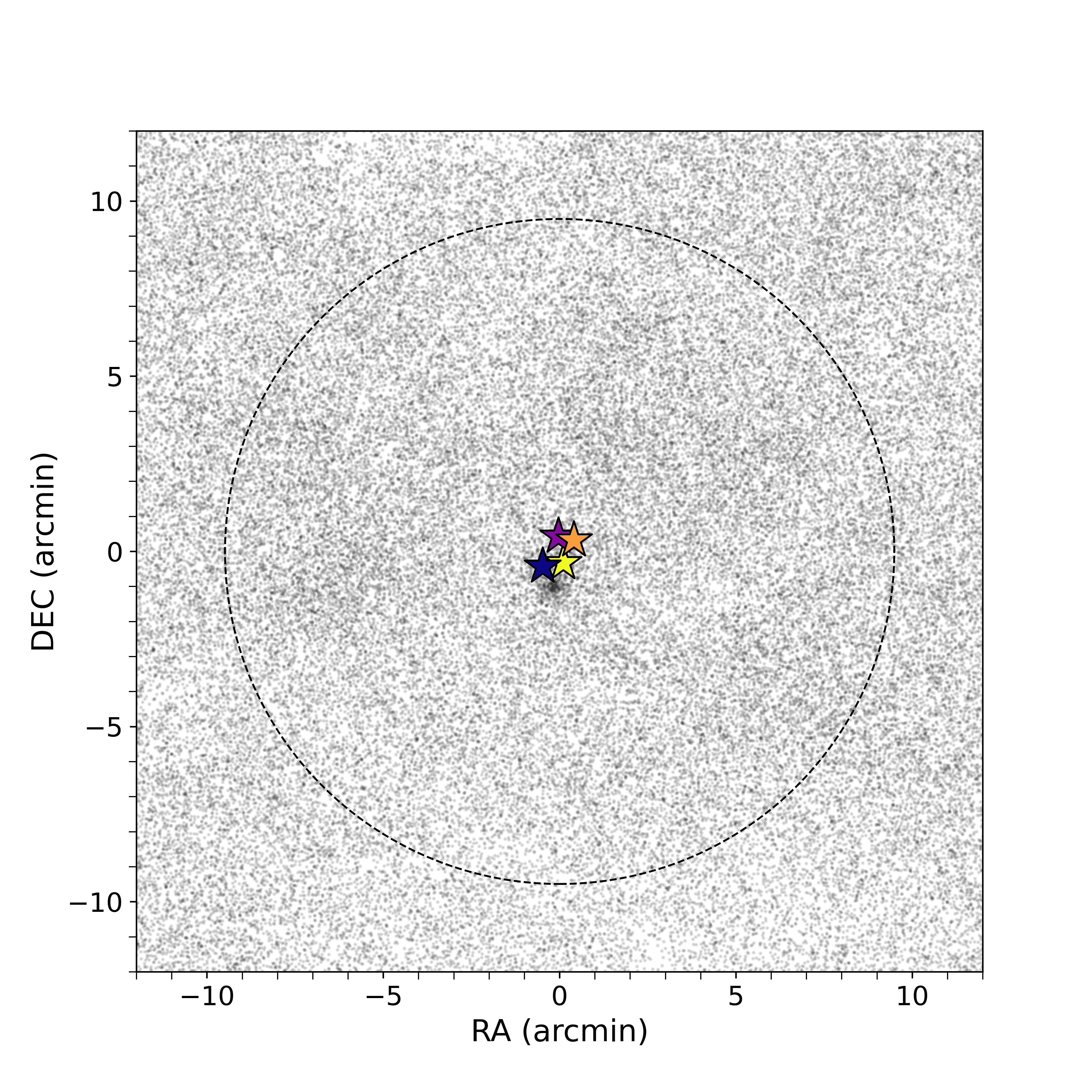}} 
    \hspace{0.3cm} 
    \subfigure[]{\includegraphics[height=0.46\textwidth]{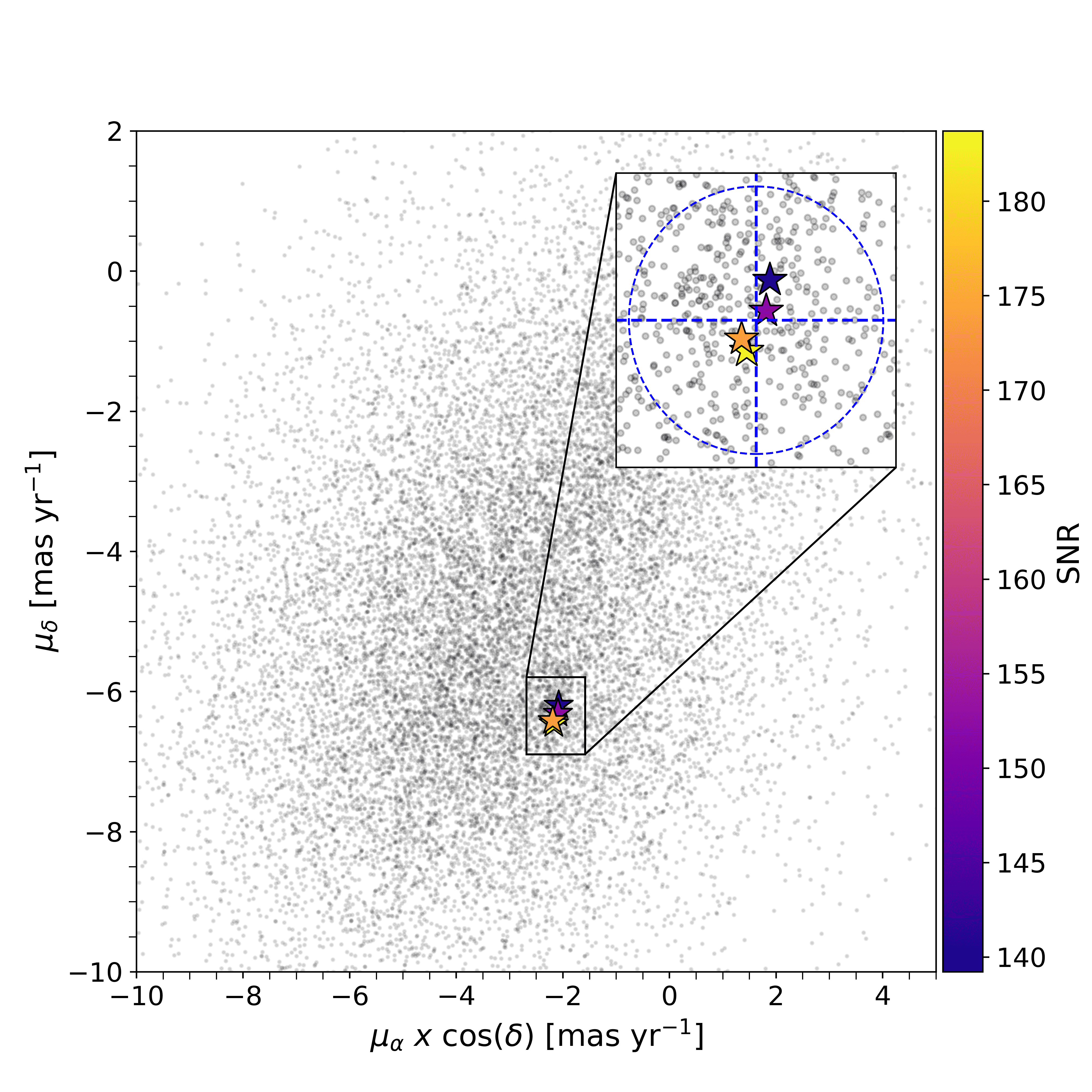}} \\
    \subfigure[]{\includegraphics[height=0.46\textwidth]{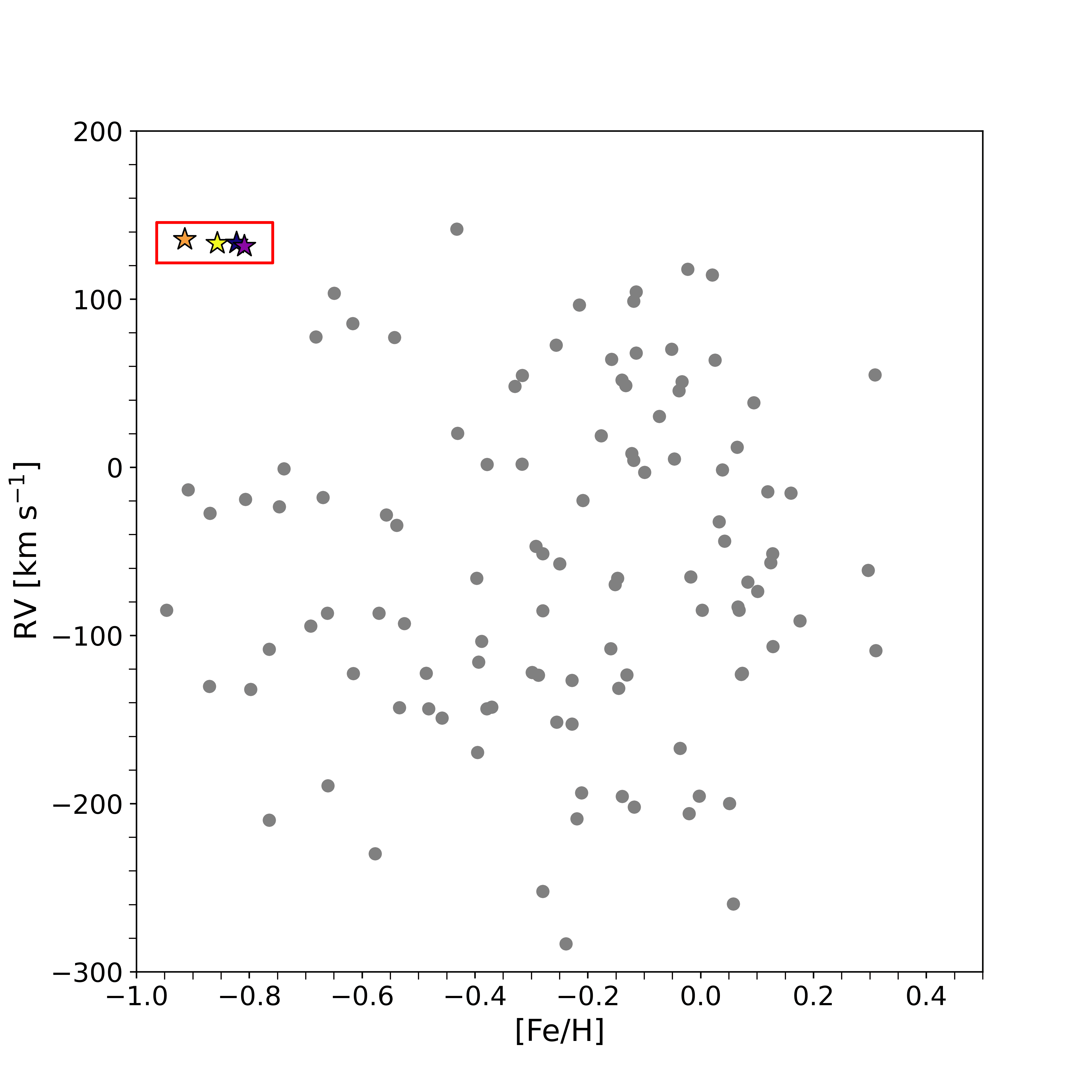}} 
    \hspace{0.3cm}
    \subfigure[]{\includegraphics[height=0.46\textwidth]{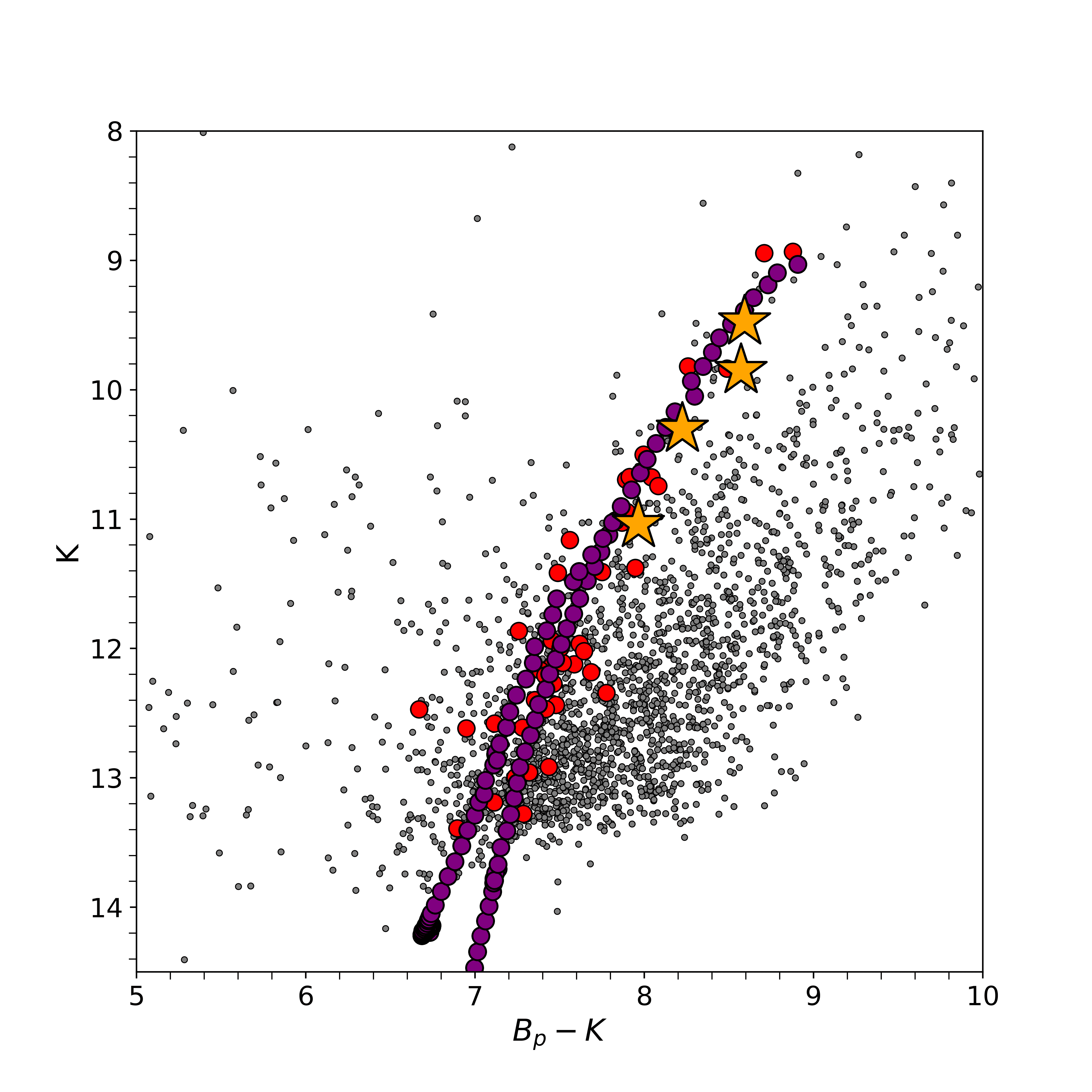}}
    \caption{\small Membership selection of Terzan 2 targets. Panels (a)-(d): (a) Spatial position. Star symbols correspond to the selected targets and are color-coded by the SNR of the APOGEE spectra (as labeled in the color-bar in the (b) panel, all panels share the same color scale). Other APOGEE targets in the field are grey dots. A circle with the tidal radius 9.49' \citep{harris10} is overplotted. (b) Proper motion density distribution of stars located within the tidal radius from the cluster center, color- and size-coded as in the (a) panel. The inner plot on the top-right shows a zoom-in of the cluster in 0.7 x 0.7 mas yr$^{-1}$ and enclosed in a proper motion radius of 0.5 mas yr$^{-1}$ shown in a blue circle. Blue dashed lines are centered on the the cluster center PM values \citep{Baumgardt2021}. (c) Radial velocity from ASPCAP versus metallicity of our members compared to field stars. The [Fe/H] of our targets have been determined with BACCHUS (see Sect. 4), while the [Fe/H] of field stars (grey dots) are taken from the ASPCAP pipeline. The astrometric, photometric and kinematic parameters used are listed in Table \ref{table1}. The red box encloses the cluster members within 0.1 dex and 10 km/s from the nominal mean [Fe/H]=$-$0.84 and RV=10.2 km s$^{-1}$ of Terzan 2, as determined in this work. (d) Colour magnitude diagram corrected by differential reddening and extinction-corrected in the Gaia Bp band and 2MASS K band of our sample and using stars within 9.49'. Our targets all lie along the red giant branch. The best isochrone fit (purple circles), corresponding to [M/H]= $-$0.80  is overplotted on Terzan 2 stars (red dots) located within 9.49 arcminutes of the cluster center and has been shifted using E(B$-$V)=2.081, A$_{v}$=5.16 and R$_{v}$=2.48.}
    \label{cuatro_imagenes}
\end{figure*}
%

\section{Target selection}
\label{sec:target selection}

The APOGEE-2S plug-plate containing Ter 2 (toward $\alpha_{J2000}=17^{h}27^{m}33.1^{s}$ and $\delta_{J2000}=-30^{d}48^{m}08.4^{s}$) was centered on $\textit{(l,b)}\sim(356.3\degree, 2.30\degree)$ inside a radious of $\leqslant$ 15' from the cluster center. The Gaia database was not available at the time of the observations, so we based our selection on existing CMDs and selecting targets within the tidal radius of the cluster \citep[9.49']{harris10}) to find the most likely members, as shown in Figure \ref{cuatro_imagenes} (a) and as described in \cite{geisler21}.

After observations were obtained, we began our search to maximize membership. The procedure is described in detail in \cite{geisler25}. Basically, we first limited our sample to observed stars within the tidal radius of the cluster \cite{Baumgardt2021}. We later matched this initial spatially selected sample with the released $\textit{Gaia}$ DR2 \citep{gaia18} catalog and required proper motions (PM)s within a radius of 0.5 mas $yr^{-1}$ around the mean PM of Ter 2: $\mu_{\alpha}\cos(\delta)=-2.13$ and $\mu_{\delta}=-6.34$ \citep{Baumgardt2021}. We initially used these values to estimate candidate members from the PMs and to reduce the potential field star contamination in the optical and NIR color–magnitude diagrams (CMDs). We later re-examined PM from the original sample using the data from the $\textit{Gaia}$ DR3 data release \citep{gaia23}. Figure \ref{cuatro_imagenes} (b) shows the PMs using Gaia DR3 \citep{gaia23}.\\
Figure \ref{cuatro_imagenes} (c) also shows the BACCHUS [Fe/H] abundance ratios (see Section \ref{sec:results}) versus the radial velocity of our four cluster members compared to field stars with ASPCAP/APOGEE-2 [Fe/H] determinations, which were shifted by $+$0.11 dex in order to minimize the systematic differences between ASPCAP/APOGEE-2 and BACCHUS, as highlighted in Appendix D of \cite{trincado20c}. We note that all of our targets have very similar velocities that are extreme compared to the field star distribution. This again supports the cluster membership for all of our targets.\\
We find a mean RV from 4 APOGEE-2 stars of 133.4$\pm$1.46 km s$^{-1}$, which is in very good agreement with the value listed in Baumgardt's web service\footnote{https://people.smp.uq.edu.au/HolgerBaumgardt/}, RV= $133.46 \pm 0.69$ km s$^{-1}$. The red box highlighted in this plot encloses the cluster members within a box centered on the mean [Fe/H]=$-$0.84 and RV=133.4 km/s  of Ter 2. After this, we were left with 4 very high probability members. We checked that all of them are positioned along the red giant branch (RGB) of the cluster, as shown in the differential reddening-corrected color magnitude diagram (see Figure \ref{cuatro_imagenes} (d)).

All selected stars had 2MASS K-band magnitudes brighter than 13. This was required in order to achieve a minimum signal-to-noise ratio (S/N) of $\gtrsim 60$ pixel$^{-1}$ in one plug-plate visit ($\sim$1 hour). The four stars observed have S/N $> 100$ pixel$^{-1}$, since more than one visit was obtained. Table \ref{table1} lists the APOGEE-2S designations of the stars, the APOGEE-2S coordinates, S/N, Gaia DR3 and 2MASS magnitudes, RV, and PMs.

\section{Atmospheric parameters}
\label{sec:atmospheric parameters}

The CMD presented in Figure \ref{cuatro_imagenes} (d) was differential reddening corrected using the same methodology as employed in \citet{rcolmenares21} and \citet{ftrincado22}. For this purpose, we selected all RGB stars within a radius of 3.5 arcmin from the cluster center and that have proper motions compatible with that of Ter 2. First, we draw a ridge line along the RGB, and for each of the selected RGB stars we calculated its distance from this line along the reddening vector. The vertical projection of this distance gives the differential interstellar absorption A$_{K}$ at the position of the star, while the horizontal position gives the differential optical+NIR reddening E(B$_{P}$-K) at the position of the star. After this first step, for each star we selected the three nearest RGB members, calculated the mean differential interstellar absorption A$_{K}$ and the mean differential reddening E(B$_{P}$-K), and finally subtracted these mean values from its B$_{P}$-K color and $K$ mangnitude. We underline the fact that the number of reference stars used for the reddening correction is a compromise between having a correction affected as little as possible by by random photometric error and the highest possible spatial resolution. 

In order to estimate the parameters of the cluster we performed isochrone fitting of the RGB using the PARSEC database \citep{bressan12}. The extinction law of \citet{cardelli89} and \citet{o'donnell94} was used. The free parametters for this fitting are the true distance modulus, (m-M)$_0$ (or the equivalent distace in pc), the interstellar absorption in the V band, A$_V$, and the reddening-law coefficient, R$_V$. These three parameters were estimated simulaneosuly using the B$_{P}$-K vs. K, B$_{P}$-R$_{P}$ vs. G and J-K vs. K CMDs, assuming an age of 12 Gyrs and initially a global metallicity of [M/H]=$-$0.8 that considers the $\alpha$-enhancement of the cluster according to the relation by \citet{salaris93}. [Fe/H] was obtained from \citet{harris10} and [$\alpha$/Fe] assumed to be +0.4. [M/H] was then adjusted during the fitting procedure in order to match the shape of the upper RGB.

The coefficient of the extinction law R$_V$ is usually assumed to be 3.1 but can vary significantly from the canonical value, especially in the direction of the Galactic Bulge \citep{nataf16} , where it can easily take lower values. In fact, we find an extinction law coefficient R$_V$ = 2.48 $\pm$ 0.1. Moreover, we achieve a fit for the distance d$_{\odot}$ = 9.8 kpc, i.e. (m-M)$_0$=14.95 $\pm$ 0.05, higher than previous estimations like 7.75$\pm$0.33 kpc from \cite{Baumgardt2021} or 7.5 kpc from \cite{harris10}, and an interstellar absorption A$_V$ = 5.16 $\pm$ 0.05. Finally, the interstellar absorption and the extinction-law coefficient we found can be translated into E(B-V)= 2.08, somewhat higher than the foreground interstellar reddening given by \citet{harris10}, i.e. E(B-V)=1.87. This discrepancy is very likely due to the higher R$_V$ used in previous studies.

The method applied to obtain T$_{eff}$ and $\log g$ from photometry was the same described in \citet{ftrincado22}. We used the differentially corrected B$_{P}$-K vs. K CMD of Figure \ref{cuatro_imagenes} (d) and projected horizontally the position of each target until it intersected the RGB of the best fitting isochrone. We then assumed the T$_{eff}$ and $\log g$ of each target to be the temperature and gravity of the point of the isochrone that has the same K$_{s}$ magnitude as the star, interpolating if necessary. We underline the fact that, for highly reddened objects like Terzan 2, the interstellar absorption correction depends on the spectral energy distribution of the star, that is, on its temperature. For this reason, we applied a temperature-dependent absorption correction to the isochrone. Without this procedure, it is not possible to obtain a proper fit of the RGB, especially of the brighter and cooler part.
Finally, with T$_{eff}$ and $\log g$ fixed, we employed the relation from \citet{mott20} for FGK stars to determine the microturbulence parameter $\xi_t$. The obtained stellar parameters are listed in \ref{table2}. 

\begin{table*}[t]
    \caption{APOGEE and Gaia DR3 data for Terzan 2 candidate stars.}
    \resizebox{1\textwidth}{!}{\begin{tabular}{c c@{\hspace{0.3cm}}c@{\hspace{0.1cm}} c@{\hspace{0.11cm}}c@{\hspace{0.1cm}} c@{\hspace{0.1cm}}c@{\hspace{0.1cm}}c@{\hspace{0.11cm}}c@{\hspace{0.1cm}}c@{\hspace{0.1cm}}c@{\hspace{0.1cm}}c@{\hspace{0.1cm}}c@{\hspace{0.1cm}}c@{\hspace{0.1cm}}c}
        \hline
        APOGEE-IDS & $\alpha$ & $\delta$ & S/N & G & BP & RP & J & H & K &RV & $\mu_{\alpha} cos(\delta)$  & $\mu_{\delta}$ \\
        &  (J2000) & (J2000)  & (pixel$^{-1}$) &  &  &  &  &  &  & (km s$^{-1}$) & (mas yr$^{-1}$) & (mas yr$^{-1}$)\\
        \hline
         2M17273185$-$3048156 & 261.89250028 & $-$30.80271932 &139&16.891&19.004&15.454&12.683&11.451&11.039 & 133.4 & $-$2.09 & $-$6.30\\
        2M17273364$-$3047243 & 261.88274825 & $-$30.80436824 &152&16.284 & 18.526 & 14.901 &  12.075 & 10.760 & 10.302 & 131.5 & $-$2.17 & $-$6.46 \\
        2M17273419$-$3048097 & 261.89019913 & $-$30.79010916 & 183& 15.768 & 18.065 & 14.329 &  11.377 & 10.019 & 9.473 & 133.2 & $-$2.08 & $-$6.19\\
        2M17273540$-$3047308 & 261.89751041 & $-$30.79190621 & 173& 16.001 & 18.419 & 14.591 & 11.636 & 10.256 & 9.848 & 135.6 & $-$2.19 & $-$6.41\\
        \hline
    \end{tabular}}
    \label{table1}
    \tablefoot{APOGEE identifications, APOGEE-2S coordinates, spectral signal-to-noise ratio (S/N), Gaia DR3 and 2MASS magnitudes, radial velocities (RVs) and proper motions (PMs) are taken from the APOGEE DR17 database.}
\end{table*}

\section{Abundance determination}
\label{sec:abundances}

The method of deriving stellar abundances is the same as that described in \citet{ftrincado22}. For this purpose we made use of the Brussels Automatic Code for Characterizing High accUracy Spectra (BACCHUS) \citet{masseron16} to derive chemical abundances for the entire sample, making a careful line selection as well as providing abundances based on a simple line-by-line approach. The BACCHUS module consists of a shell script that computes on the fly synthetic spectra for a range of abundances and compares these spectra to the observational data on a line-by-line basis, deriving abundances from different methods (e.g., using the equivalent width, the line depth, or the $\chi^{2}$). For the synthetic spectra BACCHUS relies on the radiative transfer code {\fontfamily{qcr}\selectfont Turbospectrum} (\citet{alvarez98};\citet{plez12}) and the {\fontfamily{qcr}\selectfont MARCS} model atmosphere grid \citep{gustafsson08}.

\begin{table*}[!h]
    \caption{Elemental abundances of Terzan 2 members}
    \resizebox{1\textwidth}{!}{\begin{tabular}{c c c c c c c c c c c c c c c c}
        \hline
        APOGEE ID & S/N & T$_{eff}$ & $\log g$ & $\xi_{t}$ &
        [C/Fe] & [N/Fe] & [O/Fe] & [Mg/Fe] & [Al/Fe] & [Si/Fe] & [Ca/Fe] & [Ti/Fe] & [Fe/H] & [Ni/Fe] & [Ce/Fe] \\
        \phantom{APOGEE ID} & (pixel$^{-1}$) & (K) & (cgs) & (km s$^{-1}$) &   &   &   &   &   &   &   &   &   &   & \\[-0.2ex]
        \hline
        2M17273185$-$3048156 & 139 & 4155 & 1.082 & 1.593 & -0.42 & $+$0.98 & $+$0.23 & $+$0.47 & $+$0.32 & $+$0.43 & $+$0.24 & $+$0.13 & $-$0.74 & $+$0.04 & $+$0.06 \\
        2M17273364$-$3047243 & 152 & 3975 & 0.757 & 1.654 & $-$0.31 & $+$0.96 & $+$0.25 & $+$0.41 & $+$0.31 & $+$0.27 &  $+$0.25 & $+$0.13 & $-$0.86 & $+$0.04 & $+$0.12\\
        2M17273419$-$3048097 & 183 & 3772 & 0.383 & 1.911 & $-$0.01 & $+$0.62 & $+$0.27 & $+$0.36 & $+$0.10 & $+$0.27 &  $+$0.33 & $+$0.26 & $-$0.89 & $-$0.03 & $+$0.09 \\
        2M17273540$-$3047308 & 173 & 3864 & 0.553 & 1.687 &$-$0.11 & $+$0.11 & $+$0.36 & $+$0.37 & $+$0.21 & $+$0.35 &  $+$0.30 & $+$0.26 & $-$0.88 & $-$0.03 & $+$0.15 \\
        \hline
        \rowcolor{mycolor} Mean & ... & ... & ... & ... & $-$0.21 & $+$0.81 & $+$0.28 & $+$0.40 & $+$0.24 & $+$0.33 &  $+$0.28 & $+$0.20 & $-$0.84 & $+$0.01 & $+$0.11 \\
        std & ... & ... & ... & ... & 0.16 & 0.35 & 0.05 & 0.04 & 0.09 & 0.07 & 0.04 & 0.07 & 0.06 & 0.03 & 0.03 \\
        1$\sigma$ & ...  & ... & ... & ... & 0.15 & 0.31 & 0.04 & 0.04 & 0.08 & 0.06 & 0.04 & 0.06 & 0.04 & 0.02 & 0.03 \\ 
        \hline
        \hline
        ASPCAP DR17 &  &  &  &  &  &   &   &   &   &   &   &   &   &   & \\
        \hline
        \hline
        \rowcolor{mycolor} Mean & ... & ... & ... & ... & -0.08 & $+$0.57 & $+$0.32 & $+$0.36 & $+$0.13 & $+$0.25 &  $+$0.24 & $+$0.20 & $-$0.86 & $+$0.05 & $+$0.15 \\
        std & ... & ... & ... & ... & 0.05 & 0.34 & 0.01 & 0.04 & 0.12 & 0.03 & 0.01 & 0.08 & 0.05 & 0.0 & - \\
        $1\sigma$ & ...  & ... & ... & ... & 0.04 & 0.01 & 0.03 & 0.03 & 0.08 & 0.05 & 0.01 & 0.02 & 0.03 & 0.0 & - \\ 
        \hline
    \end{tabular}}
    \tablefoot{For each chemical species, we report the mean abundance, the standard deviation and 1$\sigma$ that is defined as (84th percentil - 16th percentil)/2.} We compare our results with values obtained from ASPCAP/APOGEE DR17. Solar reference abundances are adopted from \cite{asplund05}, except for Ce, which is taken form \cite{grevesse15}
    \label{table2}
\end{table*}

The APOGEE-2 spectra provide access to 26 chemical species, on wich we were able to provide reliable abundance determinations for 11 selected chemical species, belonging to the iron-peak (Fe, Ni), odd-Z (Al), light (C,N), $\alpha$-(O, Mg, Si, Ca and Ti), and $s$-process (Ce) elements. This is due to the fact that most of the atomic and molecular lines are very weak and heavily blended, in some cases too much to produce reliable abundances in cool relatively metal-poor BGC stars. 

As described in \citet{ftrincado22} we did not include sodium in our analysis, a typical species to separate GC multiple populations, as it relies on two atomic lines (Na I: 1.6373 $\mu$m and 1.6388 $\mu$m) in the $H$ band of the APOGEE-2 spectra, which are generally very weak and heavily blended by telluric features and therefore cannot produce reliable [Na/Fe] abundance determinations in GC with the T$_{eff}$ and metallicities typical of Ter 2. This is why we focus on the elemental abundances of Al, Mg, C, N and O, which are  the chemical signatures that allow to distinguish stars with different chemical compositions in the multiple population phenomenon. \\
So from these synthetic spectra, the BACCHUS code then identifies the continuum points for normalizing the observed spectra and the relevant pixels (i.e., a mask or window) to use for the abundance determination of each line in each spectrum. The normalization is performed by selecting continuum regions in the synthetic spectra within 30 $\AA$ around the line of interest, and fitting a linear relation to these regions in the observed spectrum. Then the observed spectrum is divided by this fit.\\
The spectral window for each line is defined by identifying the regions where variations in elemental abundance cause noticeable changes in the synthetic spectra. This is done by analyzing the second derivative of the observed flux to locate the local maxima on either side of the target line.\\
For each element and spectral line, the abundance determination follows the procedure described in \cite{hawkins16}, \cite{fertrincado21}, \cite{cunha17} and \cite{rcolmenares21}; thus, we only provide a brief summary here. The abundance of each chemical species was derived through the following steps: (a) a spectral synthesis was performed using the complete atomic and molecular line list from \cite{shetrone15}, \cite{hasselquist16}, \cite{smith21}, internally labeled as \textit{linelist.20170418} (reflecting the creation date in YYYYMMDD format), which was also used to determine the local continuum level via a linear fit; (b) cosmic ray and telluric contamination were identified and removed; (c) the local signal-to-noise ratio (S/N) was estimated; (d) flux points contributing to each absorption line were automatically selected; and (e) the observed spectrum was compared with a grid of convolved synthetic spectra spanning a range of abundances to determine the best fit. Finally, four different methods were applied to derive the abundances:

\begin{enumerate}
    \item Line-profile fitting (the $\chi^{2}$ method), which determines the abundance by minimizing the squared differences between the synthetic and observed spectra. This approach fits the entire line profile, providing a comprehensive evaluation of the spectral match.
    \item Core line-intensity comparison (the int method), which focuses on matching the intensity at the core of the line between the synthetic and observed spectra. This method is particularly effective for analyzing the central features of spectral lines, where the abundance signature is most pronounced.
    \item Global goodness-of-fit estimation (the syn method), which derives the abundance by minimizing the overall discrepancy between the synthetic and observed spectra, assessing the global quality of the fit rather than focusing on specific line features.
    \item Equivalent-width comparison (the eqw method), which estimates the abundance required to reproduce the observed equivalent width of a spectral line. This widely used technique reduces the spectral information to a single value, the area under the line profile, simplifying the comparison process.
\end{enumerate} 
Each diagnostic yields validation flags. Based on these flags, a decision tree then rejects or accepts each estimate, keeping hte best-fit abundance. We adopted the $\chi^{2}$ diagnostic as the final abundance because it is the most robust \citep{hawkins16}. However, we stored the information from the other diagnostics, including the standard deviation among all four methods. \\
$^{12}$C, $^{14}$N and $^{16}$O elemental abundances are derived through a combination of vibration-rotation lines of $^{12}$C$^{16}$O  and  $^{16}$OH, along with electronic transitions of $^{12}$C$^{14}$N. The procedure for CNO analysis begins by setting the C abundance using $^{12}$C$^{16}$O lines. With this carbon abundance, the O abundance is determined using $^{16}$OH lines. If this O abundance differs from the initial value, which is scaled from the solar value by the stellar [Fe/H] ratio, the $^{12}$C$^{16}$O lines are re-analyzed  with the updated O abundance, until consistent C and O abundances are obtained from both $^{12}$C$^{16}$O and $^{16}$OH. When consistent values for C and O are obtained, the $^{12}$C$^{14}$N lines are used to derive the N abundance, which has little to no effect on the $^{12}$C$^{16}$O and $^{16}$OH lines, but the final abundances of C, N and O provide self-consistent results from $^{12}$C$^{16}$O, $^{16}$OH, $^{12}$C$^{14}$N \citet{smith13}. The molecular dependencies of CNO are extracted from this process.\\
In order to provide a consistent chemical analysis, we redetermined the chemical abundances assuming as input the effective temperature (T$_{eff}$), surface grativy ($\log g$) and metallicity ([Fe/H]) as derived by the ASCAP pipeline \citep{garciaperez16}. However, we also applied a simple approach of fixing T$_{eff}$ and $\log g$ to values determined independently of spectroscopy, in order to check for any significant deviation in chemical abundances and to minimize a number of cafeats present in ASCAP/APOGEE-2 abundances for GCs (see, e.g., \cite{masseron19}; \cite{meszaros20}; \cite{rcolmenares21}), as it is affected by a systematic effect that most likely overestimates the T$_{eff}$ values of 2P stars \citep{geisler21}.\\
The final values of each elemental abundance are summarized in Table \ref{table2}, where we compare them with the mean values estimated by ASPCAP.
We also compareed our results with other BGCs of similar metallicity like M107 from \citet{meszaros20} and NGC 6569 from \citet{barrera25}, as shown in Figure \ref{fig3}. 

The total uncertainty $\sigma_{tot}$ is calculated as the square root of the sum of the squares of the individual uncertainties obtained from the errors in each atmospheric parameter and S/N treated independently, according to the following equation:

\begin{equation}
        \sigma^{2}_{tot} = \sigma^{2}_{T_{eff}} + \sigma^{2}_{log(g)} + \sigma^{2}_{\xi_{t}} + \sigma^{2}_{S/N}
\end{equation}

where each deviation $\sigma$ represents the change in our initial central abundance estimation after running BACCHUS with atmospheric parameters varied by $T_{eff} \pm$ 100 K, $\log g \pm$ 0.3 cgs and $\xi_{t} \pm$ 0.05 km/s. These values were chosen as they represent the typical conservative uncertainties in the atmospheric parameters for our sample. This implies running BACCHUS for each of the different atmospheric parameters and statistically measuring the variations. The results of this analysis are reported in Table \ref{table4} for the star with the highest S/N in our sample, which consequently exhibits the smallest uncertainties. These uncertainties are comparable to those of the other stars in our sample.

\begin{table}[!h]
    \centering
    \caption{Typical abundance uncertainties for Terzan 2 target star}
    \label{table4}
    \resizebox{\columnwidth}{!}{ 
        \begin{tabular}{l c c c c c}
            \hline
            Abundance & $\triangle$ T$_{eff}$ & $\triangle$ $\log$g & $\triangle$$\xi$$_{t}$ & S/N & $\sigma_{Tot}$ \\
                      & (K) & (cgs) & (km s$^{-1}$) &(pixel$^{-1}$) & (dex) \\
            \hline
        {[}C/Fe{]}  & 0.01 & 0.09 & 0.02 & 0.05 & 0.10 \\
        {[}N/Fe{]}  & 0.03 & 0.10 & 0.00 & 0.06 & 0.12 \\
        {[}O/Fe{]}  & 0.05 & 0.01 & 0.04 & 0.04 & 0.08 \\
        {[}Mg/Fe{]} & 0.03 & 0.05 & 0.03 & 0.06 & 0.09 \\
        {[}Al/Fe{]} & 0.04 & 0.03 & 0.04 & 0.08 & 0.10 \\
        {[}Si/Fe{]} & 0.02 & 0.04 & 0.03 & 0.07 & 0.09 \\
        {[}Ca/Fe{]} & 0.05 & 0.06 & 0.02 & 0.03 & 0.09 \\
        {[}Ti/Fe{]} & 0.04 & 0.07 & 0.03 & 0.03 & 0.09 \\
        {[}Fe/H{]}  & 0.04 & 0.03 & 0.02 & 0.06 & 0.08 \\
        {[}Ni/Fe{]} & 0.06 & 0.02 & 0.01 & 0.04 & 0.08 \\
        {[}Ce/Fe{]} & 0.05 & 0.08 & 0.02 & 0.04 & 0.10 \\
           \hline
        \end{tabular}
    }
    \tablefoot{Typical abundance uncertainties determined for 2M172734193048097, the star with the highest S/N of our targets.}
\end{table}

\section{Results}
\label{sec:results}

\subsection{Iron-peak elements (Fe and Ni)}

\begin{figure}[h]
    \centering
    \includegraphics[width=1.0\columnwidth]{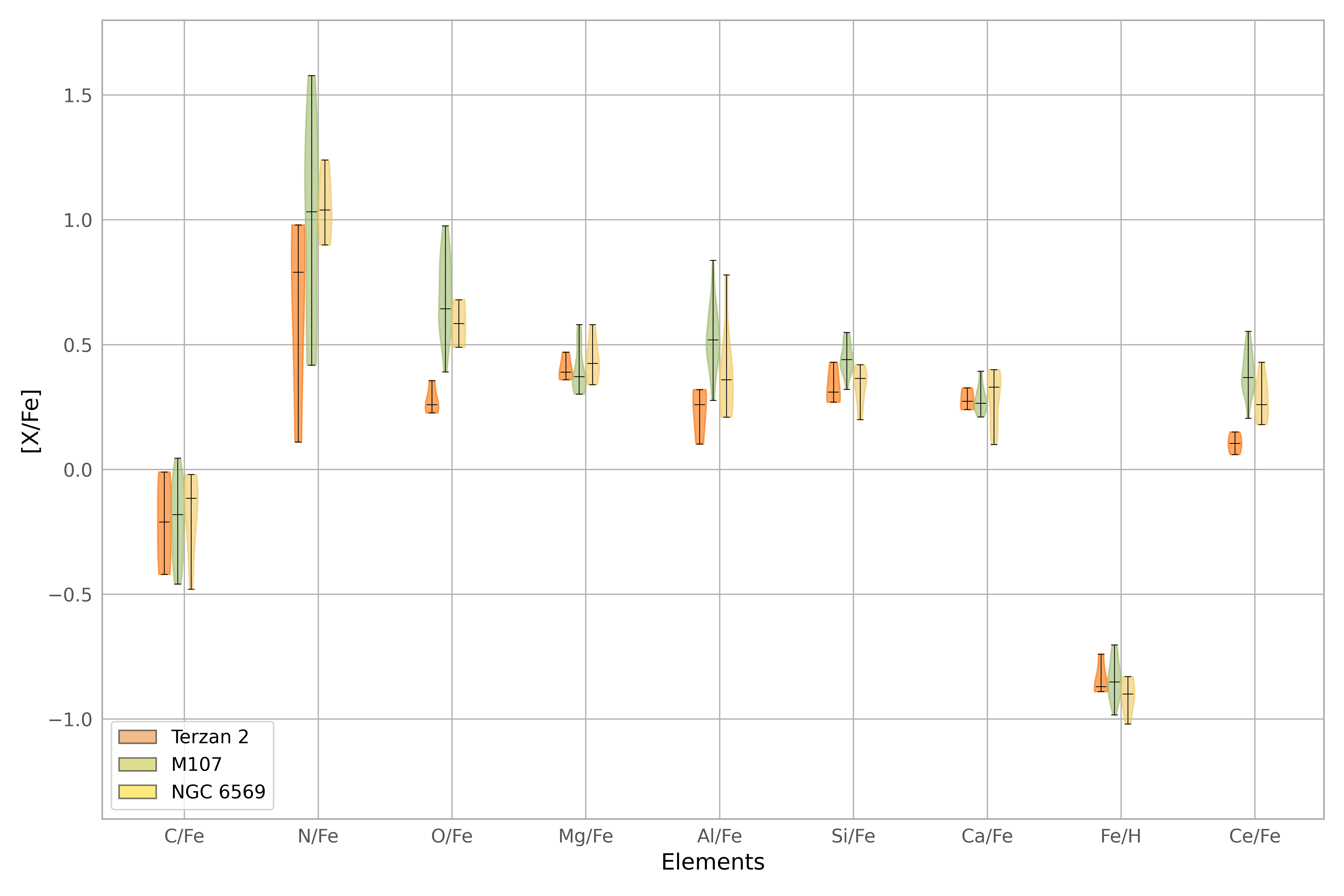} 
    \caption{[X/Fe] and [Fe/H] abundance density estimation (violin representation) of Terzan 2 (orange), compared to elemental abundances of M107 (green) from \cite{meszaros20} and NGC 6569 (yellow) form \cite{barrera25}}
    \label{fig3}
\end{figure}

Our analysis yields a mean metallicity of [Fe/H]=-0.84$\pm$0.04, with a dispersion of 0.06 dex, which is mainly produced by the relatively high metallicity ($-$0.74) of the star 2M172731853048156. However, comparing the observed dispersion with the total iron error in Table \ref{table2}, we find that there is no significant metallicity spread, which confirms that Ter 2 is consistent with other GCs with similar metallicity, as seen in Figure \ref{fig3}.

It is important to note that our mean metallicity is in very good agreement with ASPCAP DR17, as shown in Table \ref{table2}. The value of \cite{harris10} [Fe / H] for this cluster is $-$0.69, based on integrated low-resolution Ca triplet (CaT) spectroscopy \citep{armandroff88} and near-IR low-resolution spectra of seven stars \citep{stephens04}. On the other hand \citet{geisler21} gives an even closer value, [Fe/H]=-0.85$\pm$0.04 from their initial ASPCAP analysis. While
\cite{geisler25} find a value of [Fe/H]=$-$0.88$\pm$0.02 from basically the same data but using ASPCAP (but note that ASPCAP abundances are not available for 2M17273364$-$3047243 star). However, \cite{geisler23} derived a significantly higher value of [Fe/H]=$-$0.54$\pm$0.10 from Ca triplet data for eight members.

The metallicity we find for Terzan 2 places it near the the high-metallicity tail of the dominant, metal-poor peak of the BGC metallicity distribution \citep{geisler25}. BGCs of comparable metallicities have been poorly studied within a radius of $\sim$4 kpc of the Galactic center. There are some of these with little detailed chemical information (like Pal 6, Pal 12, E3). For this reason, in Fig. \ref{fig3} we compare our chemical abundances obtained for Ter 2 with M107, a BGC taken from \citet{meszaros20}, and NGC 6569 from \cite{barrera25}, a BGC from CAPOS that have similar metallicity to our cluster. In this figure, we do not plot Ti and Ni since \citet{meszaros20} does not report their abundances.

An additional iron-peak element considered in our analysis is nickel (Ni). We obtained an average of <[Ni/Fe]=$+$0.01>$\pm$0.02, with a dispersion of 0.03 dex. This abundance is slightly lower ($\lesssim$0.04 dex) than the abundance ratios measured by {\fontfamily{qcr}\selectfont ASPCAP DR17}. The abundance ratio [Ni/Fe] in Ter 2 is similar to that observed in objects of similar metallicity, as seen in Figure \ref{fig3}.

\begin{figure*}[h]
    \centering
    \includegraphics[width=1.0\textwidth]{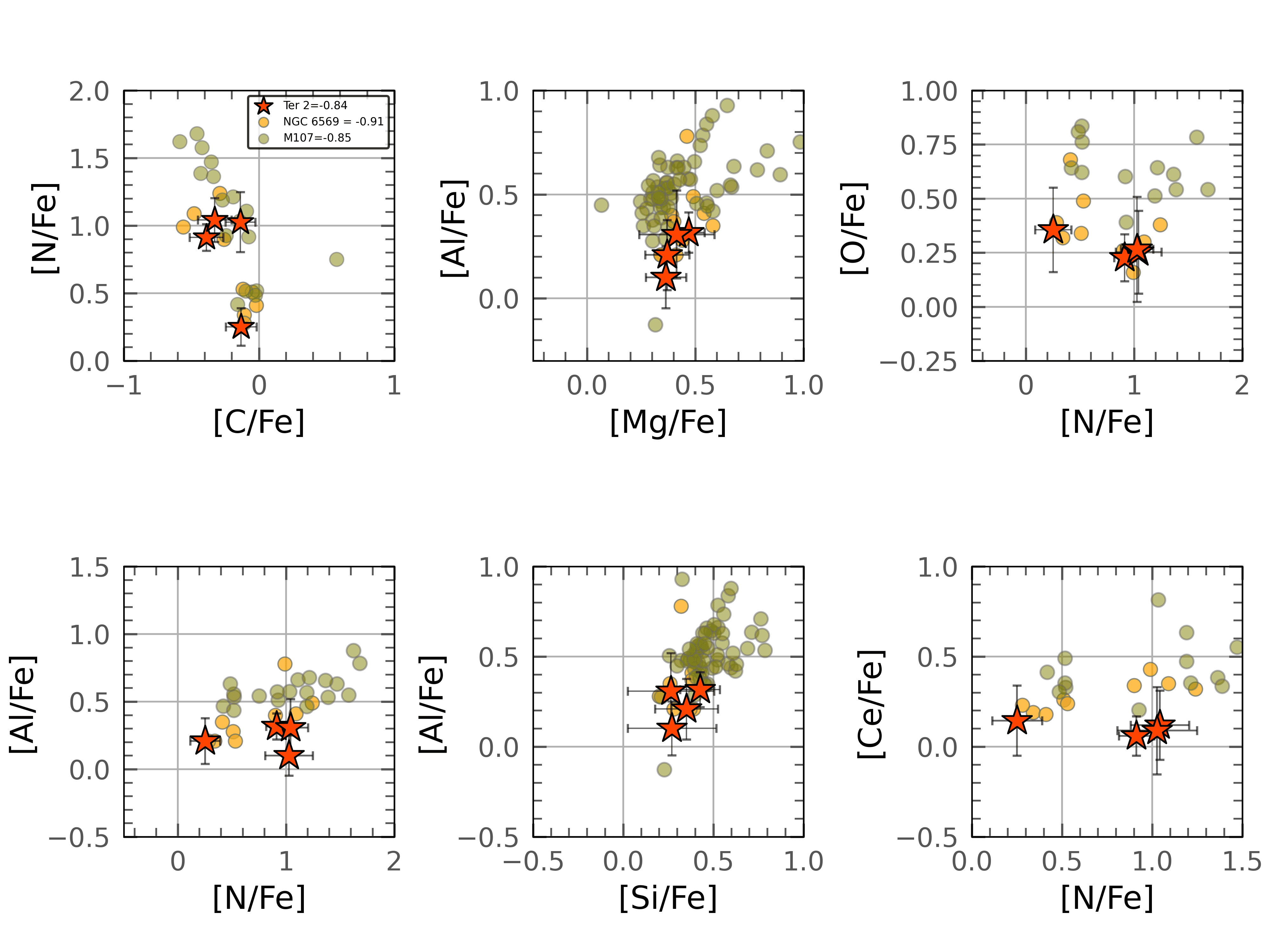} 
    \caption{From left to right, the top row shows the distributions of [N/Fe]-[C/Fe], [Al/Fe]-[Mg/Fe], [O/Fe]-[N/Fe]; the bottom row shows [Al/Fe]-[N/Fe], [Al/Fe]-[Si/Fe] and [Ce/Fe]-[N/Fe]. Data correspond to Terzan 2 stars from this work (orange stars), M107 (green circles) from \citet{meszaros20}, and NGC 6569 (yellow circles) from \citet{barrera25}. Typical uncertainties for Terzan 2 stars are also indicated.}
    \label{fig4}
\end{figure*}

\subsection{The light elements C and N}
We find that Ter 2 exhibits a high enrichment in nitrogen, with a mean abundance of [N/Fe]=$+$0.81$\pm$0.31, with a large  dispersion of 0.35 dex, and an average carbon deficiency of [C/Fe]=$-$0.21$\pm$0.15 with a dispersion of 0.16 dex, which is lower than the value [C/Fe]=-0.08 calculated by ASPCAP DR17, while our [N/Fe] value is higher.

Figure \ref{fig4} shows a clear anti-correlation in C-N as well as N-O.  However, there is no sign of an N-Al correlation. The behavior of C and N is comparable to that observed in M107 and NGC 6569. The lack of a N-Al correlation in Ter 2 could be due to our limited sample of targets and future observations are required to verify this finding. Also, measurements indicate that Ter 2 is populated by a fraction of stars with enhanced [N/Fe] abundances well above typical galactic values ([N/Fe]$\gtrsim$ $+$0.5). This is a common behavior for GCs, and a clear indication of multiple stellar populations (see, e.g.,\citet{schiavon17};\citet{ftrincado2020b};\citet{geisler21}). Interestingly, one of the analyzed stars, 2M17273540$-$3047308, shows a rather low [N/Fe] abundance, more consistent with first stellar population. This contrast further reinforces the presence of a chemical spread in nitrogen and is compatible with the existence of multiple populations in Terzan 2.

\begin{figure*}[h!]
    \centering
    \includegraphics[width=1.0\textwidth]{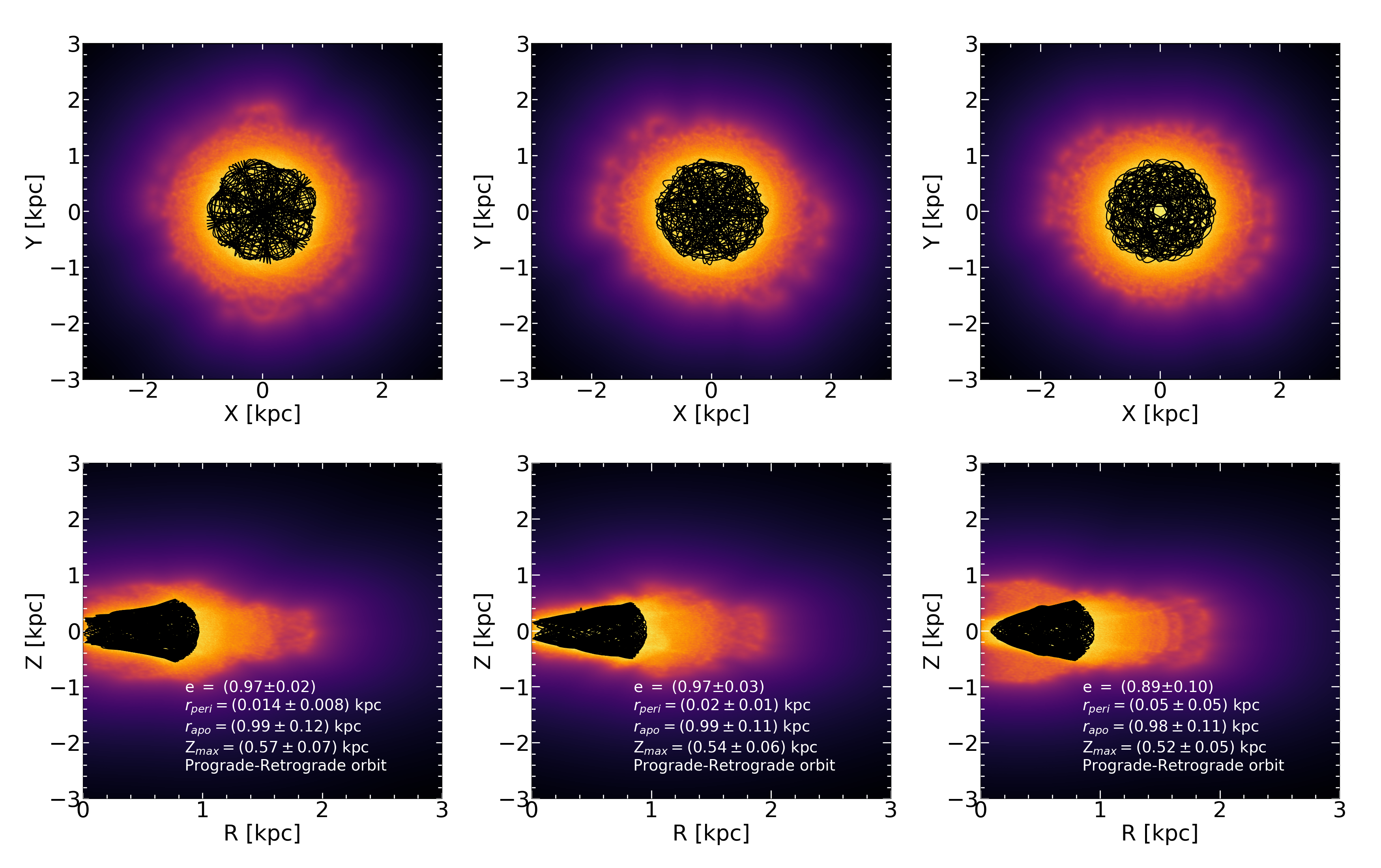}  
    \caption{Ensemble of one million orbits for Terzan 2 in the frame corotating with the MW bar, projected on the equatorial $(top)$ and meridional $(bottom)$ Galactic planes in the non-inertial reference frame with a bar pattern speed of 31 $(left )$, 41 $(middle)$ and 51 $(right)$ km s$^{-1}$ kpc$^{-1}$, and time-integrated backward over 2 Gyr. The yellow and orange colors correspond to more probable regions of the space, which are most frequently crossed by the simulated orbits. The black solid line shows the orbital path of Terzan 2 from the observables without error bars.}
    \label{fig5}
\end{figure*}

\subsection{The $\alpha$-elements: O, Mg, Si, Ca and Ti}

Figure \ref{fig3} and Table \ref{table2} show that Ter 2 exhibits a considerable $\alpha$-element enhancement, with mean values of: [O/Fe]=$+$0.28$\pm$0.04, [Mg/Fe]=$+$0.40$\pm$0.04, [Si/Fe]=$+$0.33$\pm$0.06, [Ca/Fe]=$+$0.28$\pm$0.04 and [Ti/Fe]=$+$0.20$\pm$0.06, with no evidence of an intrinsic spread in any of these elements. These values are in good agreement with GCs of similar metallicity.
The $\alpha$-elements in Ter 2 are overabundant compared to the Sun, which is a common behavior in Galactic GCs and field stars at similar metallicity, such as M107 and NGC 6569. The agreement with ASPCAP DR17 is generally very good, with a difference of 0.02 dex in the worst case (that is, for [Si / Fe]).\\
\cite{geisler25} find the same mean [Si/Fe] from ASPCAP. They also point out a systematic difference in [Si/Fe] for the CAPOS clusters in general between the BACCHUS and ASPCAP abundances, with BACCHUS on average 0.17$\pm$0.07 dex higher.

\subsection{The odd-Z element Al and the s-process element Ce}
Ter 2 exhibits a mean aluminum enrichment of [Al/Fe]=+0.24$\pm$0.08, with a dispersion of 0.09 dex. We did not find a  variation in [Al/Fe] beyond the typical errors. However, if we compare [Al/Fe] vs. [Mg/Fe] (see Fig. \ref{fig4}), we see that there is a hint of Mg-Al. If real, this means that the Mg-Al cycle barely started in the stars that were responsible for the intracluster contamination. This is consistent with expectations of AGB nucleosynthesis (\citet{ftrincado21d}; \citet{ventura08}; \citet{ventura16}; \citet{karakas14}). 

We find a mean abundance of [Ce/Fe]=+0.11$\pm$0.03, which is slightly super-Solar, and comparable to the Ce levels observed in other Galactic GCs at similar metallicity (see, e.g., \citet{masseron19}; \citet{meszaros20}). The observed spread for the $s$-process element Ce is $+$0.03 dex, with no indication of intrinsic dispersion.  Figure \ref{fig4} reveals that there is no correlation between Ce and N. This result does not show the correlation between Ce and N observed by \cite{ftrincado22} in the clusters Ton 2 and \cite{ftrincado21} in NGC 6380, suggesting that the chemical enrichment history of Terzan 2 may differ from that of those clusters.

The agreement with ASPCAP DR17 is very good for [Ce/Fe] with a difference of 0.04 dex, and slightly worse for [Al/Fe], with a difference of 0.11 dex.

\section{Dynamical history}
\label{sec:dynamical history}

We made use of the state-of-art Milky Way model --\texttt{GravPot16}\footnote{\url{https://gravpot.utinam.cnrs.fr}} to predict the orbital path of Terzan 2 in a steady-state Galactic gravitational model that includes a ``boxy/peanut" bar structure \citep{ftrincado19}. Orbits have been integrated with the \texttt{GravPot16}, which includes the perturbations due to a realistic (as far as possible) rotating "boxy/peanut" bar, which fits the structural and dynamical parameters of the Galaxy based on recent knowledge of our Milky Way. \\
For the orbit computations, we adopted the same model configuration, solar position and velocity vector as described in \citet{ftrincado19}, except for the angular velocity of the bar $\Omega_{bar}$, for which we employed the recommended value of 41 km s$^{-1}$ kpc$^{-1}$ \citep{Sanders2019} and assuming variations of $\pm$ 10 km s$^{-1}$ kpc$^{-1}$. The considered  structural parameters of our bar model (e.g., mass and orientation) are within observational estimations that lie in the range of 1.1$\times$10$^{10}$M$_{\odot}$ and present-day orientation of 20$\degree$ (value adopted from dynamical constraints of \cite{Tang_2018}) in the non-inertial frame (where the bar is at rest). The bar scale lengths are $x_{\circ}$=1.46 kpc, $y_{\circ}$=0.49 kpc, and $z_{\circ}$=0.39 kpc, and the middle region ends at the effective semi-major axis of the bar R$_{c}$=3.28 kpc \citep{robin12}.\\
For guidance, the Galactic convention adopted by this work is: $X$-axis is oriented toward $l$=0$\degree$ and $b$=0$\degree$, $Y$-axis is oriented toward $l$=90$\degree$; the velocity is also oriented in these directions. Following this convention, the Sun's orbital velocity vectors are [U$_{\odot}$, V$_{\odot}$, W$_{\odot}$]=[11.1, 12.24, 7.25] km s$_{-1}$ \citep{schonrich10}. The model has been rescaled to the Sun's Galactocentric distance, 8.27 kpc \citep{gravity21}, and a circular velocity at the solar position to be $\sim$ 229 km s$^{-1}$ \citep{eilers19}. Table \ref{tab:orbit_params} gives our input parameters and their errors.\\
The most likely orbital parameters and their uncertainties were estimated using a simple Monte Carlo scheme. An ensemble of one million orbits was computed backward in time for 2 Gyr, under variations of the observational parameters assuming a normal distribution for the uncertainties of the input parameters (e.g., positions, heliocentric distances, radial velocities, and proper motions), which were propagated as 1$\sigma$ variations in a Gaussian Monte Carlo resampling. To compute the orbits, we adopt a mean radial velocity from the member stars from APOGEE DR17 data \citep{abdurro22} as reported by \citet{Baumgardt2021}. The nominal proper motions ($<\mu_{RA}>$ and $<\mu_{DEC}>$) for Terzan 2 were taken from Gaia EDR3  \citep{gaia21}, assuming an uncertainty of 0.5 mas yr$^{-1}$ for the orbit computations. The heliocentric distance ($d_{\odot}$) was also adopted from \citet{Baumgardt2021}. The results for the main orbital elements are listed in the inset of Figure \ref{fig5}. 

\begin{table*}[t]
    \centering
    \caption{Input parameters adopted for the orbital integration of Terzan 2.}
    \label{tab:orbit_params}
    \resizebox{0.85\textwidth}{!}{
        \begin{tabular}{lcc}
            \hline
            Parameter & Value & Reference \\
            \hline
            Radial velocity (RV) & $+133.46 \pm 0.69$ km s$^{-1}$ & \citet{Baumgardt2021} \\
           Proper motion in RA ($\mu_{\alpha}\cos{\delta}$) & $-2.141 \pm 0.03$ mas yr$^{-1}$ & \citet{gaia21} \\
           Proper motion in Dec ($\mu_{\delta}$) & $-6.255 \pm 0.03$ mas yr$^{-1}$ & \citet{gaia21} \\
            Heliocentric distance ($d_{\odot}$) & $7.8 \pm 0.3$ kpc & \citet{Baumgardt2021} \\
           Bar pattern speed ($\Omega_{\mathrm{bar}}$) & $41 \pm 10$ km s$^{-1}$ kpc$^{-1}$ & \citet{Sanders2019} \\
           Solar position ($R_\odot$, $Z_\odot$) & 8.2 kpc, 25 pc & \citet{ftrincado2020b} \\
          Solar velocity vector & ($U$, $V$, $W$) = (11.1, 245.8, 7.8) km s$^{-1}$ & \citet{ftrincado2020b} \\
           \hline
       \end{tabular}
    }
\end{table*}
Figure \ref{fig5} shows the simulated orbital path of Terzan 2 by adopting a simple Monte Carlo approach. The probability densities of the resulting orbits are projected on the equatorial and meridional galactic planes, in the non-inertial reference frame where the bar is at rest. The black line in the figure shows the orbital path (adopting observables without uncertainties). The yellow color corresponds to the most probable regions of the space, which are crossed more frequently by the simulated orbits. Terzan 2 is clearly confined in a bulge-like orbit, which lies in an in-plane orbit with high eccentricity $>$0.9 and low vertical (Z$_{max}$ $<$ 0.6 kpc) excursions from the Galactic plane, and apogalactocentric distances below 1 kpc. This cluster is going inside and outside of the bar in the Galactic plane,but not with a bar-shaped orbit, which means that this cluster is not trapped by the bar, but exhibits a chaotic orbital motion which experiences both prograde and retrograde orbits, whose effect is produced by the presence of the bar. From the average orbital parameters listed in Figure \ref{fig5}, we can see that the Galactic bar induced the orbit to have smaller average (peri-/apo-) galactic distances, lower vertical excursions from the Galactic plane, and higher eccentricities, even for any of the bar pattern speed values, suggesting that the presence of different $\Omega_{bar}$ in Terzan 2 are almost negligible.\\
The bulge nature of this cluster derived from its orbit here agrees with the classification by \cite{geisler25} based on a synthesis of a number of both dynamical as well as chemical analyses.




\section{Conclusions}
\label{sec:conclusions}
We present the first detailed elemental-abundance analysis for four stars of the Bulge Globular Cluster Terzan 2, as part of the CAPOS survey using APOGEE-2 data. We examine 11 chemical species: the light elements C and N, the $\alpha$-elements O, Mg, Si, Ca, and Ti, the iron-peak elements Fe and Ni, the odd-Z element Al, and the s-process element Ce using the BACCHUS code. Overall, the chemical species examined so far in Ter 2 are in general agreement with other Galactic GCs at similiar metallicity from \citet{meszaros20} and \cite{barrera25}. The main conclusions of this paper are the following:
   \begin{itemize}
      \item The mean metallicity of Terzan 2 is [Fe/H]=$-$0.84$\pm$0.04, with a star-to-star [Fe/H] spread which is not statistically significant, indicating a homogeneous iron content. 
      \item Our value of metallicity is lower than the classical value of $-$0.69 from \cite{harris10}. It is in good agreement with other ASPCAP studies of the same cluster, in particular $-$0.88$\pm$0.02 from \cite{geisler25}. However, it is significantly lower than values derived using the CaT technique, such as $-$0.54$\pm$0.10 from \cite{geisler23}, pointing to the difficulty of obtaining accurate metallicities for such highly obscured GCs and such small samples. It is a member of the dominant lower-metallicity peak of the bulge GC metallicity distribution.
      \item Even with the low number of stars analyzed, there is an apparent anticorrelation between C-N, a common feature among globular clusters (e.g. \cite{meszaros20}). A spread in [N/Fe] has been detected in Terzan 2 ($\sim$0.5 dex). This result indicates that Terzan 2 is  likely to host the multiple-population phenomenon, but any firm conclusion should await a substantial increase in sample size.
      \item The alpha elements align with the trends commonly observed in globular clusters within the same metallicity range \citep{meszaros20} with an enhancement of between 0.2 $-$ 0.4 dex.
      \item We did not find evidence for a real spread in Ce abundances in Terzan 2. Furthermore, the apparent Ce–N and Ce–Al correlations reported in other bulge globular clusters (e.g., Ton 1 and Ton 2) at similar metallicity (see Fernández-Trincado et al. 2021, 2022) are not observed in our data.
      \item Our dynamical analysis suggests that Terzan 2 is a Bulge GC. It lives in the inner bulge region, but goes inside and outside of the bar in the Galactic plane but with a non-bar-shaped orbit, which means that Terzan 2 is not trapped by the bar, but exhibits a chaotic orbital motion which describes prograde and retrograde paths at the same time, whose effect is produced by the presence of the bar.
   \end{itemize}

\begin{acknowledgements}
      Part of this work was supported by the German \emph{Deut\-sche For\-schungs\-ge\-mein\-schaft, DFG\/} project number Ts~17/2--1. S.V. gratefully acknowledges the support provided by Fondecyt Regular n. 1220264 and by the ANID BASAL project FB210003. D.G. gratefully acknowledges the support provided by Fondecyt Regular n. 1220264.\\
      D.G. also acknowledges financial support from the Direcci\'on de Investigaci\'on y Desarrollo de la Universidad de La Serena through the Programa de Incentivo a la Investigaci\'on de Acad\'emicos (PIA-DIDULS).
\end{acknowledgements}

%
%

\bibliographystyle{aa} 
\bibliography{aa} 

\end{document}